\documentclass[prl,showpacs,twocolumn]{revtex4}
\usepackage{graphics}
\usepackage{epsfig}
\usepackage{graphicx}

\begin{document}
\title{Entangled multi-qubit states without higher-tangle}
\author{Heng Fan$^1$, Yong-Cheng Ou$^1$, Vwani
Roychowdhury$^2$
 }
\affiliation{$^1$Institute of Physics,
Chinese Academy of Sciences, Beijing 100080, China\\
$^2$Electrical Engineering Department, University of California
Los Angeles, Los Angeles, CA 90095, USA}

\date{\today}

\begin{abstract}
We analyze mixed multi-qubit states composed of a $W$ class state
and a product state with all qubit in $|0\rangle $. We find the
optimal pure state decomposition and convex roofs for
higher-tangle with bipartite partition between one qubit and the
rest qubits for those mixed states. The optimality of the
decomposition is ensured by the Coffman-Kundu-Wootters (CKW)
inequality which describes the monogamy of quantum entanglement.
The generalized monogamy inequality is found to be true for $W$
class states with arbitrary partitions between one qubit and
multi-qubit.
\end{abstract}

\maketitle

Quantum entanglement has been the subject of much study in recent
years as a physical resource for quantum communication and quantum
information processing \cite{horodecki}. Entangled states have a
number of remarkable features which has inspired an enormous
literature in the years since their discovery. As a consequence,
the study of quantum entanglement from various view points has
been a very active area and has led to many interesting results.
However, only the pure state entanglement shared between two
parties is thoroughly understood and quantified; progress on
mixed state of higher-dimension and the multipartite state has
been much slower. In this paper, we will study the higher-tangle
of mixed multiqubit states and the related monogamy of
entanglement.

Monogamy of entanglement is a key property discovered recently in
the context of multiqubit entanglement \cite{CKW,T}. It states
that unlike classical correlations, quantum entanglement cannot
be freely shared among many parties. For example, if a pair of
qubits Alice ($A$) and Bob ($B$) have perfect quantum
correlation, namely, if they are maximally entangled, then Alice
cannot be entangled to a third party Charlie ($C$). The monogamy
inequality is to capture in a quantitative way the tradeoff
between quantum entanglement shared by pair $(A,B)$ and by pair
$(A,C)$. In the context of quantum cryptography \cite{E91}, such a
monogamy property is of fundamental importance since it
quantifies how much information an eavesdropper could potentially
obtain about the secret key to be extracted. The triqubit
monogamy inequality was first proposed and proved by Coffman,
Kundu and Wootters (CKW) in their seminal paper \cite{CKW}, and
it is also named as CKW inequality. Recently the long standing
conjecture of the general monogamy inequality with bipartite
partition between one qubit and the rest qubits for multiqubit
states is proved by Osborne and Verstraete \cite{OV}. To be
precise, a multiqubit state $\rho _{AB_1...B_n}$ shared among
$n+1$ parties, the distribution of bipartite entanglement
satisfies the monogamy inequality (also refer as CKW inequality):
\begin{eqnarray}
\tau (\rho _{AB_1})+\tau (\rho _{AB_2})+...+\tau (\rho _{AB_n})
\le \tau (\rho _{A:B_1B_2...B_n}), \label{monogamy}
\end{eqnarray}
where the bipartite quantum entanglement is measured by {\it
tangle} $\tau $ which is the square of the well known {\it
concurrence} \cite{HW,W}. For $\tau (\rho _{A:B_1B_2...B_n})$, the
entanglement with bipartite partition for multiqubit is across
$A:B_1B_2...B_n$ cut. It is obvious from this monogamy inequality
that the summation of quantum entanglement measured by tangle in
pairwise type $\tau (\rho _{AB_1}),...$, and $\tau (\rho
_{A,B_n})$ is upper bounded by the amount of entanglement with
bipartite partition between $A$ and $B_1...B_n$ measured by $\tau
(\rho _{A:B_1B_2...B_n})$.

In the monogamy inequality (\ref{monogamy}), the bipartite
entanglement is measured by tangle $\tau $ which is the square of
the concurrence for pure states, as we mentioned. The {\it
concurrence} introduced in Refs.\cite{HW,W} is directly related
with entanglement of formation  \cite{BDSW} for two-qubit case.
As we know, the entanglement of formation is a well accepted
entanglement measure, and has a clear physical implication in
quantum communication protocols such as quantum teleportation
\cite{teleportation}. However, in general, the {\it concurrence},
also with entanglement of formation can only be explicitly
calculated for pure states and two-qubit mixed states. The reason
is that the {\it concurrence} and the entanglement of formation of
a mixed state are represented by the convex roof of the pure state
decompositions \cite{U}. As we know, there is no general method
to find the optimal pure state decomposition for a mixed state.
However, recently, Lohmayer {\it et al} \cite{LOSU} successfully
find the optimal decomposition of mixed three-qubit states
composed of a GHZ state and a $W$ state. And the optimal
decomposition and the convex roofs for the three-tangle $\tau _3$
\cite{CKW} are obtained, where three-tangle for three-qubit state
$\rho _{ABC}$ is defined as $\tau _3(\rho _{ABC})\equiv \tau
(\rho _{A:BC})-\tau (\rho _{AB})-\tau (\rho _{AB})$, the residual
entanglement between $A$ and $BC$ that cannot be accounted for by
the entanglement of $A$ with $B$ and $C$ separately. Similarly,
we refer $\tau (\rho _{A:B_1...B_n})-\sum _{j=1}^n\tau (\rho
_{AB_j})$ to higher-tangle. The monogamy inequalities by other
entanglement measures and related topics 
can be found in Refs.\cite{M,KW,CW1,O,OF,DVC}

In this paper, we shall start from the analyzing of the tangle
for $(n+1)$-qubit {\it mixed} states composed of a $W$ class state
and the state $|0\rangle ^{\otimes (n+1)}$. The optimal
decomposition for this kind of mixed states is found and the
optimality is ensured by the CKW inequality. We then consider a
more general monogamy inequality for a $W$ class state with
arbitrary partitions which is beyond the scope of the case proved
in Ref.\cite{OV}. Really, by the obtained result of tangle of the
mixed state under consideration, the monogamy inequality is found
to be still true for this case. And the monogamy inequalities by
other entanglement measures will be presented. The study of this
topic may shed light into both the entanglement measure for
multipartite mixed states, a well known problem in entanglement
theory of quantum information processing, and the monogamy in
quantum entanglement distribution. More references about those
problems can be found in a nice recent review paper of quantum
entanglement in Ref.\cite{horodecki}.

{\it Tangle of multiqubit mixed states.}--Let's consider a
$(n+1)$-qubit $W$ class state defined as:
\begin{eqnarray}
|W\rangle &=&a|100...0\rangle +b_1|010...0\rangle
+b_2|001...0\rangle +... \nonumber \\
&&...+b_n|000...1\rangle , \label{wstate}
\end{eqnarray}
where generally $a, b_j, j=1,...,n$ are complex numbers. Without
lose of generality, we assume that they are real numbers. This
assumption does not change any of our conclusions in this paper.
And as usual, we have the normalization condition $a^2+\sum
_{j=1}^nb_j^2=1$. Suppose we have a $(n+1)$-qubit mixed state
shared by parties $A$ and $B_1$, $B_2$,..., and $B_n$ which is a
mixture of the state $|W\rangle $ and the state $|\vec{0}\rangle
\equiv |0\rangle ^{\otimes (n+1)}$,
\begin{eqnarray}
\rho _{AB_1...B_n}=p|W\rangle \langle W|+(1-p)|\vec{0} \rangle
\langle \vec{0}|, \label{mixed}
\end{eqnarray}
where $p$ is the probability of the $W$ class state. We would
like to find the bipartite state tangle $\tau (\rho
_{A:B_1B_2...B_n})$ between $A$ and $B_1B_2...B_n$.

Before proceed, let's introduce the definition of the concurrence.
For a pure bipartite state $|\phi _{AB}\rangle $, the concurrence
is defined as ${\cal {C}}(\phi _{AB})=2\sqrt{{\rm det}\rho _A}$,
where ${\rm det}\rho _A$ is the determinant of matrix $\rho _A$,
$\rho _A={\rm tr}_B\left( \phi _{AB}\right) $ is the reduced
density operator of $\phi _{AB}$,  and ${\rm tr}_B$ is to take
trace over Hilbert space $B$. Here we use the notation $\phi
_{AB}=|\phi _{AB}\rangle \langle \phi _{AB}|$. The concurrence of
the mixed state $\rho _{AB}$ is defined as the average pure state
concurrence minimized over all pure state decompositions,
\begin{eqnarray}
{\cal {C}}(\rho _{AB})={\rm min}\sum p_jC(\phi ^j_{AB}),
\end{eqnarray}
which is also called convex roof extension \cite{U}, where $\rho
_{AB}=\sum p_j\phi ^j_{AB}$. The optimal decomposition means that
the pure state decomposition which gives the defined mixed state
concurrence. The optimal decomposition is not necessarily unique.
For two-qubit mixed state, the concurrence can be found by a well
known method introduced by Wootters \cite{W}. However, for the
general mixed state of multi-qubit, there does not exist an
explicit formula for the concurrence.

Tangle of a pure state is defined as the squared concurrence.
Similarly tangle of a mixed state is defined as the average pure
state tangle over all pure state decompositions,
\begin{eqnarray}
\tau (\rho _{AB})={\rm min}\sum p_j\tau (\phi ^j_{AB})= {\rm
min}\sum p_j{\cal {C}}^2(\phi ^j_{AB}). \label{tangle}
\end{eqnarray}
We next will study the tangle of the mixed state in (\ref{mixed}),
$\tau (\rho _{A:B_1...B_n})$. We remark that though $\rho
_{AB_1...B_n}$ is a multiqubit state, the tangle $\tau (\rho
_{A:B_1...B_n})$ is for a bipartite partition between $A$ and
$B_1...B_n$.

Stimulated by the work in Ref.\cite{LOSU}, see \footnote{ In
Ref.\cite{LOSU}, the authors study the triqubit mixed state
composed of $W$ state $(|001\rangle +|010\rangle +|100\rangle
)/\sqrt{3}$ and the GHZ state $(|000\rangle +|111\rangle )/\sqrt
{2}$. This case is different from our case in several aspects:
Our case is a multiqubit state; We use the $W$ class state which
has different amplitude in each term compared with the $W$ state;
We use product state of $|0\rangle $ instead of the GHZ state in
the mixture; The final conclusion is different}, we propose a
trial pure state decomposition for the mixed state in
(\ref{mixed}) as the following:
\begin{eqnarray}
\rho _{AB_1...B_n}=\frac {r}{3}\sum _{k=0}^2|\psi ^k\rangle
\langle \psi ^k|+(1-r)|\vec{0}\rangle \langle \vec{0}|,
\label{trialmixed}
\end{eqnarray}
where $|\psi ^k\rangle =\sqrt {q}|W\rangle +\sqrt {1-q}\omega ^k
|\vec{0}\rangle $, and $\omega =e^{\frac {2\pi ik}{3}}$. Compare
this pure state decomposition with the state in (\ref{mixed}), we
find that equation $rq=p$ should be satisfied. For pure state
$|\psi ^k\rangle $,  the reduced density operator of qubit $A$,
$\rho _A^k={\rm tr}_{B_1...B_n}\psi ^k$, takes the form
\begin{eqnarray}
\rho _A^k&=& (\sqrt{q}a|1\rangle +\sqrt {1-q}\omega ^k|0\rangle
)(\sqrt{q}a|1\rangle +\sqrt {1-q}\omega ^k|0\rangle )^{\dagger }
\nonumber \\
&&+q\sum _jb_j^2|0\rangle \langle 0|.
\end{eqnarray}
By the equation $\tau (\psi ^k)=4{\rm det}\rho _A^k$, the tangle
of the pure state $|\psi ^k\rangle $ with bipartite partition
across $A:B_1...B_n$ cut can be calculated as $\tau (\psi
^k)=4q^2a^2(\sum _jb_j^2)$.

By definition we know that tangle $\tau (\rho _{A:B_1...B_n})$
should be less than or equal to the average tangle in the trial
pure state decomposition (\ref{trialmixed}),
\begin{eqnarray}
\tau (\rho _{A:B_1...B_n})&\le &\frac {r}{3}\times \sum
_{k=0}^2\tau (\psi ^k) \nonumber
\\
&=&4rq^2a^2(\sum _jb_j^2) \nonumber \\
&=&4pqa^2(\sum _jb_j^2),
\end{eqnarray}
where the last equation is due to the condition $rq=p$, and also
when $r=1$, $q$ is minimum $q=p$, we thus know $\tau (\rho
_{A:B_1...B_n}) \le 4p^2a^2(\sum _jb_j^2)$. We next will show that
this is a tight upper bound for tangle, e.g., the trial pure state
decompositions can realize the optimal pure state decomposition.

According to the CKW inequality (\ref{monogamy}) proved in
Ref.\cite{OV} , we know that $\tau (\rho _{A:B_1...B_n})\ge \sum
_{j=1}^n\tau (\rho _{AB_j})\ge \sum _{j=1}^n{\cal {C}}^2 (\rho
_{AB_j})$, where the last inequality is because that the squared
concurrence is a convex function on the set of density matrices
\cite{CKW}, it is further shown to be an inequality in Ref.\cite{osborne} .
With the help of Eq.(\ref{mixed}), we obtain
\begin{eqnarray}
\rho_{AB_j}&=&p\left( a|10\rangle +b_j|01\rangle \right) \left(
a|10\rangle +b_j|01\rangle \right) ^{\dagger } \nonumber \\
&&+[1-p+\sum _{k\not =j}b_k^2] |00\rangle \langle 00|.
\end{eqnarray}
The squared concurrence can be calculated as ${\cal {C}}^2(\rho
_{AB_j})=4a^2b_j^2p^2$ by the formula in Ref.\cite{W}. So we find
the lower bound of the tangle has the form $\tau (\rho
_{A:B_1...B_n}) \ge 4p^2a^2(\sum _jb_j^2)$. Combine the upper
bound and lower bound, we have a conclusion that
\begin{eqnarray}
\tau (\rho _{A:B_1...B_n}) &=& 4p^2a^2(\sum _jb_j^2) \nonumber \\
&=&\sum _{j=1}^n\tau (\rho _{AB_j})\label{tight}.
\end{eqnarray}
We thus know that the optimal pure state decomposition of tangle
for state $\rho _{A:B_1...B_n}$ can be in the form presented in
Eq.(\ref{trialmixed}). Actually only three vectors are enough to
realize the optimal pure state decomposition since $r=1$ is for
all region of $p$, where $p\in [0,1]$. Since the monogamy
inequality (\ref{monogamy}) is tight for the case considered in
this paper as shown in (\ref{tight}), the higher-tangle defined
as $\tau (\rho _{A:B_1...B_n})-\sum _{j=1}^n\tau (\rho _{AB_j})$
vanishes for the state in (\ref{mixed}), while this mixed state
is in general entangled except in some extreme cases.

{\it Monogamy inequality with arbitrary partitions for $W$ class
states.}--A nature question arise concerning about the CKW
inequality in (\ref{monogamy}) is whether this monogamy inequality
is generally true or not for higher-dimensional systems. A simple
example in Ref.\cite{O} shows that this monogamy inequality does
not hold in general for higher-level quantum states. Still we may
wonder, to what extent, this monogamy inequality can be applied,
for example, even for case of multiqubit states. We next consider
a following question: A multiqubit state $\rho _{ABCD...}$ shared
by $A,B,C,D$,..., etc., while particle in $A$ is a qubit, but
$B,C,D,...,$ contains several qubits as $B=(B_1B_2...B_n)$,
$C=(C_1C_2...C_m)$, $D=(D_1D_2...D_l)$,.... From the CKW
inequality we know that both $\tau (\rho _{A:BCD...})$ and $\tau
(\rho _{A:B})+\tau (\rho _{A:C})+\tau (\rho _{A:D})+...$, are
greater than or equal to quantity $\sum _{j_B}\tau (\rho
_{AB_{j_B}})+\sum _{j_C}\tau (\rho _{AC_{j_C}})+\sum _{j_D}\tau
(\rho _{AD_{j_D}})+...$. The problem is for arbitrary partitions,
whether we still have the following monogamy inequality?
\begin{eqnarray}
 \tau (\rho
_{A:BCD...})\ge  \tau (\rho _{A:B})+\tau (\rho _{A:C})+\tau (\rho
_{A:D})+... \label{general}
\end{eqnarray}

We may consider a normalized $W$ class state shared by $ABCD...$
as the follows,
\begin{eqnarray}
|\tilde {W}\rangle &=&\tilde {a}|100...0\rangle +[\sum
_{j_B}\tilde {b}_{j_B}\sigma ^x_{j_B}+\sum _{j_C}\tilde
{c}_{j_C}\sigma ^x_{j_C} \nonumber \\
&&+\sum _{j_D}\tilde {d}_{j_D}\sigma ^x_{j_D}+...]|\vec{0}\rangle
, \label{wstate1}
\end{eqnarray}
where $\sigma ^x_k$ is the Pauli matrix on site $k$. This state
is actually just like the state in (\ref{wstate}). By direct
calculation and as pointed out in \cite{CKW} , we know the $W$
class states saturate the CKW inequality. That means $\tau (\rho
_{A:BCD...})=\sum _{j_B}\tau (\rho _{AB_{j_B}})+\sum _{j_C}\tau
(\rho _{AC_{j_C}})+\sum _{j_D}\tau (\rho _{AD_{j_D}})+...$ for
state $|\tilde {W}\rangle $ in Eq. (\ref{wstate1}). Then it is
quite interesting to know whether all of the states $\rho _{AB},
\rho _{AC}, \rho _{AD},...$, saturate the CKW inequality or not
since we already know $\tau (\rho _{A:BCD...})\le  \tau (\rho
_{A:B})+\tau (\rho _{A:C})+\tau (\rho _{A:D})+...$. The
saturation means that this is an equation, otherwise, the CKW
inequality for partition $\rho _{A:BCD}, \rho _{A:B}, \rho
_{A:C}, ...$ will be violated. Thus $W$ class states in
(\ref{wstate1}) are good candidates to check the CKW inequality
since there is no room for other than saturation case, otherwise,
a counterexample is found to violate the CKW inequality.

Without lose of generality, let's study the reduced multiqubit
mixed state $\rho _{AB}$, from Eq.(\ref{wstate1}), we find
\begin{eqnarray}
\rho _{AB}&=&(\tilde {a}|10...0\rangle +\tilde {b}_1|01...0\rangle
+...+\tilde {b}_n|00...1\rangle )\times \nonumber \\
&&(\tilde {a}|10...0\rangle +\tilde {b}_1|01...0\rangle
+...+\tilde {b}_n|00...1\rangle )^{\dagger } \nonumber \\
&&+(\sum _{j_C}\tilde{c}^2_{j_C}+\sum_{j_D}\tilde{d}^2_{j_D}+...)
|0...0\rangle \langle 0...0|.
\end{eqnarray}
Let $p=\tilde {a}^2+\sum _{j}\tilde {b}_j^2$, and let $a=\tilde
{a}/\sqrt {p}, b_j=\tilde {b}_j/\sqrt {p}$, quite interestingly,
we find $\rho _{AB}= \rho _{AB_1...B_n}$, as presented in
(\ref{mixed}). Thus, all mixed states $\rho _{AB}$, $\rho _{AC}$,
$\rho _{AD}$,..., are exactly the same form as the mixed states in
(\ref{mixed}).

We already find the optimal pure state decomposition of the
multiqubit mixed state $\rho _{AB_1...B_2}$, and really it
saturates the CKW inequality, e.g., without higher-tangle. We thus
know that all mixed states $\rho _{AB}$, $\rho _{AC}$, $\rho
_{AD}$,..., saturate the CKW inequality. So for $W$ class states,
the generalized CKW inequality (\ref{general}) still holds for
bipartite case with one party to be a qubit and other parties can
be in arbitrary partitions $B,C,D...$. Interestingly, the CKW
inequality is saturated for all of those partitions. One may
realize now that not only $W$ class states, actually all states
(mixed) in from (\ref{mixed}) actually saturate the CKW
inequality with arbitrary partitions, see Figure. Here, we are
tempted to conjecture that for all multiqubit systems, a more
general monogamy inequality as (\ref{general}) holds. This
question of course should be explored further.

\begin{figure}
  \centering
\includegraphics[width=2.5in]{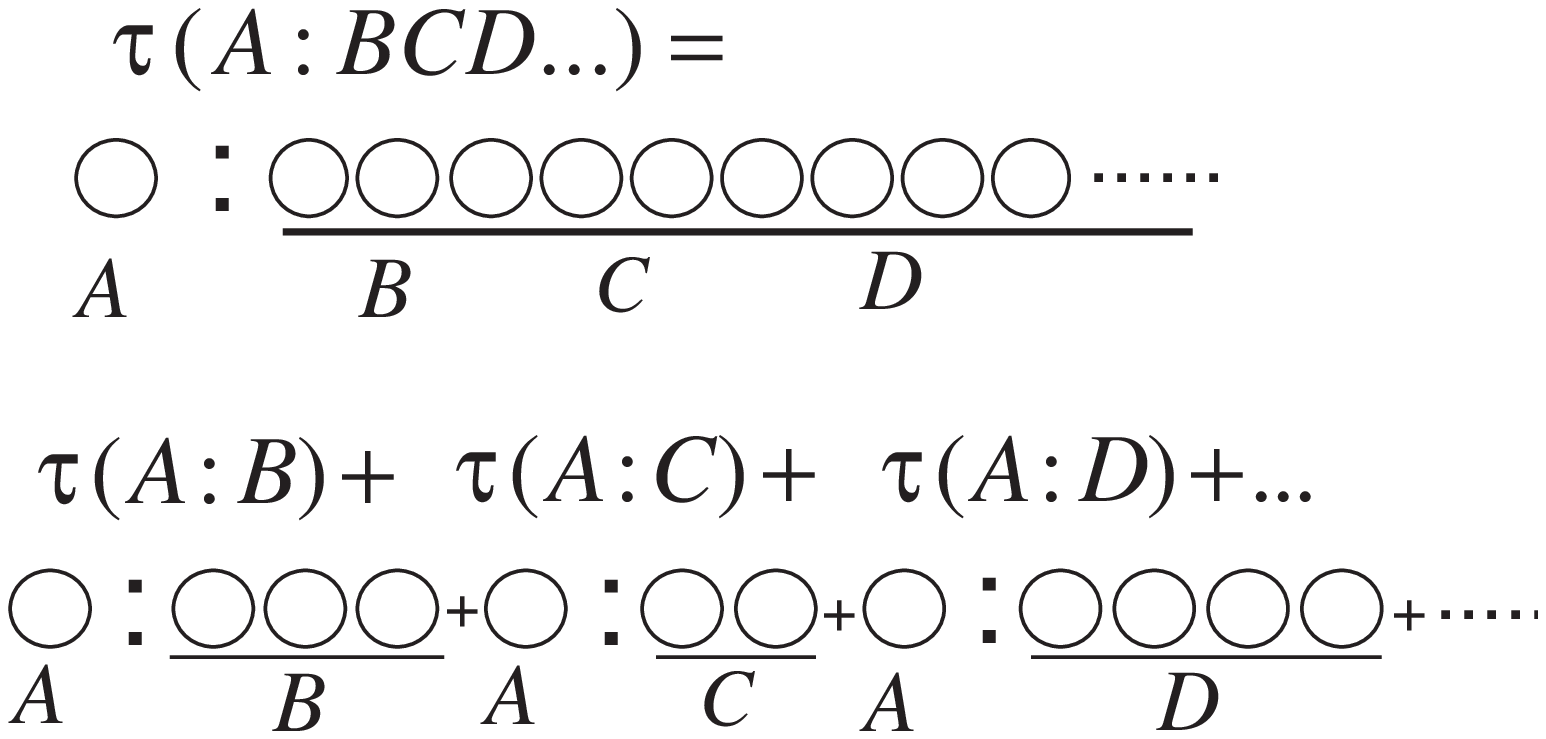}
  \caption{A generalized CKW inequality is satisfied (saturated) for mixed states in (\ref{mixed}) form,
  if $p=1$, they reduce to $W$ class states.}\label{figure}
\end{figure}

{\it An example.}--Let's consider a five-qubit $W$ state,
$|W_{ABC}\rangle =(|10000\rangle +|01000\rangle
+...+|00001\rangle ) /\sqrt{5}$, where $B=B_1B_2$ and $C=C_1C_2$.
Since $W$ state is symmetric, and we have $\rho _{AB}=\rho
_{AC}$. Our motivation is to find whether we have a general
monogamy inequality as $\tau (\rho _{A:BC})\ge \tau (\rho
_{A:B_1B_2})+\tau (\rho _{A:C_1C_2})$. One can find $\tau (\rho
_{A:BC})=4{\rm det}\rho _A=\frac {16}{25}$. Since $W$ state
saturates the CKW inequality, then if $\tau (\rho _{A:B})\not
=\frac {8}{25}$, see \footnote{Apparently it should be strictly
larger than $8/25$ since of the CKW inequality in
(\ref{monogamy}), with this we have $\tau (\rho _{A:B_1B_2})\ge
2\tau (\rho _{AB_1})=\frac {8}{25}$.}, it is a counter-example
that violates the general CKW inequality. For $W$ state, we have
$\rho _{AB_1B_2}=\frac {3}{5}|W'\rangle \langle W'|+\frac
{2}{5}|000\rangle \langle 000|$, where we denote $|W'\rangle
\equiv (|100\rangle +|010\rangle +|001\rangle )/\sqrt {3}$. With
our conclusions in this paper, the optimal pure state
decomposition takes the form $\rho _{A:B_1B_2}=\frac {1}{3}\sum
_{k=0}^2|{\psi '}^k\rangle \langle {\psi '}^k|$, where $|{\psi
'}^k\rangle =\sqrt {\frac {3}{5}}|W'\rangle +\sqrt {\frac
{2}{5}}\omega ^k|000\rangle $. Then by definition in
(\ref{tangle}), we have $\tau (\rho _{A:B_1B_2})=\frac {8}{25}$.
Thus for $W$ state, a more general monogamy inequality is
satisfied.

{\it Monogamy inequality of entanglement by other entanglement
measures.}-- Three-tangle $\tau _3$ is defined due to the CKW
inequality \cite{CKW,M} as the residual entanglement. Since it
vanishes for any separable pure state, it is then proposed to
define the genuine multi-qubit entanglement \cite{DVC}. However,
$W$ states ($W$ class states) which are genuine multipartite
entangled have vanishing three-tangle, as we already know. We
thus propose to use other monogamy inequalities in terms of
different entanglement measures to quantify the residual
entanglement. In Ref.\cite{OF}, the monogamy inequality by
entanglement measure, negativity denoted as ${\cal {N}}(\rho
_{AB})$ \cite{VW}, is proposed. It is shown that the residual
entanglement is always larger than zero for $W$ like states.
Similar as in Ref.\cite{OF}, we would like to point out that, the
entanglement measure by the realignment method \cite{CW}, see
also \cite{fan}, denoted as ${\cal {R}}(\rho _{AB})$, also
satisfies the monogamy inequality. To be precise, let $|\psi
_{ABC}\rangle $ be a triqubit state, then we have
\begin{eqnarray}
{\cal {R}}^2(\psi _{A:BC})&\ge &{\cal {R}}^2(\rho _{AB})+{\cal
{R}}^2(\rho _{AC}),
\end{eqnarray}
similar as the monogamy inequality in terms of the negativity
${\cal {N}}^2(\psi _{A:BC})\ge {\cal {N}}^2(\rho _{AB})+{\cal
{N}}^2(\rho _{AC})$. Those two inequalities are due to the fact
that, ${\cal {R}}(\rho _{AB})$ and ${\cal {N}}(\rho _{AB})$ are
lower bounds for concurrence ${\cal {C}}(\rho _{AB})$ as pointed
out in Ref.\cite{CAF}, ${\cal {C}}\ge {\rm max} \{ {\cal
{N}},{\cal {R}}\} $. Thus the proof of CKW inequality leads to
the proof of those two inequalities. Those two inequalities can be
complementary to each other. We remark that one advantages to use
${\cal {N}},{\cal {R}}(\rho _{AB})$ is that they are operational
to be calculated. The monogamy inequality by relative entropy of
entanglement \cite{VPRK,V} is also an 
interesting question \cite{FP}.

{\it Conclusions.}--We analyze the multiqubit mixed states
composed of a $W$ class state and the state $|\vec{0}\rangle $,
the optimal pure state decomposition is found for tangle for those
states. It is shown that these mixed states saturate the CKW
inequality and thus are without higher-tangle. We then study a
more general CKW inequality for multiqubit state with arbitrary
partitions, $W$ class states are likely to violate this general
CKW inequality. However, we find the general CKW inequality with
arbitrary partitions is still true for $W$ class states, the
general CKW inequality is saturated by $W$ class states. A new
monogamy inequality by matrix realignment quantity is presented.

{\it Acknowlegements.} H.F. was supported in part by `Bairen'
program, `973' program (2006CB921107) and NSFC. He would like to
thank S.J.Gu and H.Q.Lin for their hospitality when he was
visiting the Chinese University of Hong Kong. Y.C.O. was supported
in part by the Postdoctoral Science Foundation of China. V.R. was
partly supported by the U. S. Army Research Office/DARPA under
contract/grant number DAAD 19-00-01-0172, and in part by the NSF
grant CCF:0432296.

\end{document}